\documentclass[10pt,a4paper]{IEEEtran}
\usepackage{epsfig}
\usepackage{ifthen}
\usepackage{law}
\usepackage{mystyle}
\usepackage{graphics}
\usepackage{times}
\usepackage{cite}

\newcommand{\Draft}{0}   

\ifthenelse{\Draft = 1}
{
  \newcommand{\fixMe}[2][]{
    \typeout{***** ERROR: fixMe still in final version *****}
  }
}
{
  \newcommand{\fixMe}[2][] {[{\bf #1}] {\bf \marginpar{\large FIX}} {\em #2}}

}

\title{Canonical Polygon Queries on the Plane: A New Approach
\thanks{This is the extended and correct version of that one presented in proceedings of Australasian Workshop On Combinatorial Algorithms 2004(AWOCA 04).}
}
\author{
S. Sioutas, D. Sofotassios, K. Tsichlas, D. Sotiropoulos, P. Vlamos  \\
\normalsize Ionian University-Department of Informatics, Computer Technology Institute, Aristotle University of Thessaloniki-Informatics Department, Greece \\
Email: sioutas@ionio.gr, sofos@cti.gr, tsichlas@csd.auth.gr, \{dgs,vlamos\}@ionio.gr}
  
\date{}
\begin{document}

\maketitle
\thispagestyle{empty}

\begin{abstract}The polygon retrieval problem on points is the problem of preprocessing a set of $n$ points on the plane, so that given a polygon query, the subset of points lying inside it can be reported efficiently. It is of great interest in areas such as Computer Graphics, CAD applications, Spatial Databases and GIS developing tasks. In this paper we study the problem of canonical $k$-vertex polygon queries on the plane. A canonical $k$-vertex polygon query always meets the following specific property: a point retrieval query can be transformed into a linear number (with respect to the number of vertices) of point retrievals for orthogonal objects such as rectangles and triangles (throughout this work we call a triangle orthogonal iff two of its edges are axis-parallel). We present two new algorithms for this problem. The first one requires $O(n\log^2{n})$ space and $O(k\frac{log^3n}{loglogn}+A)$ query time. A simple modification scheme on first algorithm lead us to a second solution, which consumes $O(n^2)$ space and $O(k \frac{logn}{loglogn}+A)$ query time, where $A$ denotes the size of the answer and $k$ is the number of vertices. The best previous solution for the general polygon retrieval problem uses $O(n^2)$ space and answers a query in $O(k\log{n}+A)$ time, where $k$ is the number of vertices. It is also very complicated and difficult to be implemented in a standard imperative programming language such as C or C++.
\end{abstract}


\section{Introduction}

Given a set $S$ of $n$ points on the plane, the problem of retrieving a subset $S' \in S$ that lie in the interior of a planar geometric object is of great interest in the areas of Computational Geometry, Spatial Databases, Computer Graphics, CAD and GIS applications. The efficiency of the solutions presented so far depends on the existence or not of orthogonality on the query object, which means that not all the line segments forming the query figure are vertical or horizontal.

A range tree \cite{S90} for a example, is powerful enough to support windowing of points (i.e. the query object is an arbitrary axis-parallel rectangle) in $O(\log{n}+A)$ time using $O(n\log{n})$ space. The problem becomes harder as the complexity of the query object increases (i.e. triangle, quadrilateral, arbitrary polygon), and there is no full orthogonality.

It is important to notice that the point retrieval problem for simple polygons is an interesting problem for many application areas. In medicine for example, the term ROI, which stands for Region Of Interest, is widely used by physisians in order to indicate a polygonal region of arbitrary complexity on their scene (i.e. Radiology Image). Many information systems developed so far support retrieval queries at such a region by first computing its bounding rectangle. Then, they report all the points inside the rectangle and finally these points are filtered so that only the points inside the region remain.

Willard \cite{W82}, was the first to present a solution for the point retrieval problem for simple $k$-vertex polygons. It uses $O(n)$ space and the query time is $O(n^{0.77}+A)$. Edelsbruner and Welzl \cite{EW86}, reduced the query time to $O(n^{0.69}+A)$. A faster algorithm was presented by Edelsbruner, Kirkpatrick and Maurer \cite{EKM82} that uses $O(k\log{n}+A)$ query time and $O(n^7)$ space. Cole and Yap \cite{CY83}, presented a method using $O(n^2/\log{n})$ space and $O(k\log{n}\log{\log{n}}+A)$ time. Finally, Paterson and Yao \cite{PY86} have presented the best known solution (for a simple arbitrary polygon). This solution uses $O(n^2)$ space and answers a query in $O(k\log{n}+A)$ time.
 
In this work we consider the polygon retrieval problem on points in special case of canonical polygons which always satisfy the following strict property: a retrieval query on a set of points can be transformed to a linear number (on the number of their vertices) of queries on orthogonal objects such as rectangles and triangles. We call a triangle orthogonal iff two of its edges are axis-parallel. We present two solutions: The first one requires $O(n\log^2{n})$ space and $O(k\frac{log^3n}{loglogn}+A)$ query time. The second one consumes $O(n\log{n})$ space and $O(k \frac{logn}{loglogn}+A)$ query time, where $A$ denotes the size of the answer and $k$ is the number of vertices.

The best previous solution for the general polygon retrieval problem uses $O(n^2)$ space and answers a query in $O(k\log{n}+A)$ time, where $k$ is the number of vertices. 

This paper is organized as follows. In Section~\ref{sec:prel} we briefly introduce some preliminary data structures. In Section~\ref{sec:alg} we give the details of our algorithms. In Section~\ref{sec:thoughts} we present some special extensions for the general polygon retrieval problem based on algorithm of Paterson and Yao for which we shortly introduce the fundamental notions. Finally, some conclusions and further extensions concerning this problem are considered in Section~\ref{sec:concl}.\\

\section{Preliminary Data Structures} \label{sec:prel}

\subsection{Fusion Trees}

At ACM STOC 1990, Fredman and Willard \cite{FW90} surpassed the comparison-based lower bounds for sorting and searching using the features in a standard imperative programming languages such as $C$. Their key result was an $O(logn/loglogn)$ time bound for deterministic searching in linear space. The time bounds for dynamic searching include both searching and updates. Since then much effort has been spent on finding the inherent complexity of fundamental searching problems.

\subsection{Planar Point Location in Sublogarithmic Time}

Timothy M. Chan \cite{C06} and Mihai Patrascou \cite{P06} extended the 1-dimensional fusion tree \cite{FW90} to 2-dimensions,
in order to handle the point location problem in $O(logn/loglogn)$ time and linear $O(n)$ space.

\subsection{Half Plane Range Query}

The half plane range query problem is the problem of reporting all the points in a set $S$ of $n$ points on the plane that lie on a given side of a query line $L$. This section combines the method presented by Chazelle, Guibas, and Lee \cite{CGL83} that achieves $O(\log{n}+A)$ query time and linear space (using the notion of duality) with the best current method for planar point location problem \cite{C06}, \cite{P06}. The main steps of algorithm \cite{CGL83} are the following:

\begin{itemize}

\item Preprocessing
	
\begin{enumerate}
	
	\item Partition $S$ into a set of convex layers:\\
	(a) Define $S_i$ as the convex hull of all the points currently in $S$\\
	(b) Remove the vertices of $S_i$ from $S$\\
	(c) Increment $i$, repeat the process\\
	
	The time cost is $O(n\log{n})$ while the space complexity is $O(n)$, using a technique that computes convex hulls in a dynamic environment \cite{C83}.
	
	\item Augment the set of layers building vertical connections as follows: for each vertex $w$ of layer $S_i$, keep a pointer to the two edges immediately above and below $w$. This clearly uses $O(n)$ extra space.
	
	\item Using duality, the transformation of each vertex $w$ into its corresponding line maps each layer into another convex polygon. The produced mapping is organized into a point location structure, occupying $O(n)$ space. \label{item:point}
	
\end{enumerate}

\item Query Processing
	
\begin{enumerate}

	\item Given a query line $L$ transform it into its corresponding dual point $P_L$.
	
	\item Apply a planar point location algorithm for the point $P_L$ in the properly organized structure. This determines the innermost layer among the layers containing the point $P_L$. Thus, in the dual mapping it determines the innermost layer among the layers that $L$ intersects. Call this layer \emph{neighboring}. Using the best current point location algorithm, this costs $O(logn/loglogn)$ time.
	
	\item Using the pointers mentioned at step \ref{item:point} of the preprocessing procedure, it is easy to report one vertex lying at the query half plane for each layer which encloses the neighboring one.
	
	\item Traverse each layer from each of the vertices reported across the part of the layer inside the half plane. Report the vertices traversed. 

\end{enumerate}

\end{itemize}

Clearly, these steps lead to an algorithm for answering half plane range queries using $O(n)$ space and $O(logn/loglogn+A)$ query time. We use this method in order to answer orthogonal triangle range queries on points.

\subsection{Priority Search Tree}

In this subsection, we briefly review the priority search tree of McCreight \cite{McC85}. Let $S$ be a set of $n$ points on the plane. We want to store them in a data structure, so that the points that lie inside a semi-infinite strip of the form $(-\infty , b] \times (-\infty , c]$, can be found efficiently. 

 The priority search tree is a binary search tree over the $x$-coordinates of the points. The root of the tree contains the point $p$ with the minimum $y$-coordinate. The left (resp. right) subtree is recursively defined for the set of points in $S-\{p\}$. The set $S-\{p\}$ is partitioned equally into the two subtrees of the root. As a result, it is easy to see, that a point is stored in a node on the search path from the root to the leaf containing its $x$-coordinate.
 
\begin{figure}[htbp]
	\begin{center}
		\includegraphics[scale=0.7]{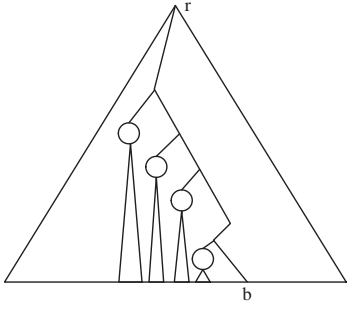}
	\end{center}
 \caption{$P_{b}$: the search path, $L_{b}$: the nodes that are left sons of nodes on $P_{b}$ and do not belong to the path.}
 \label{fig:search_path}
\end{figure}

 Queries with ranges that are half-infinite in both $x$ and $y$ directions are also known as {\em quadrant range search}. To answer a quadrant range query, we find the $O(\log{n})$ nodes in the search path $P_{b}$ for point $b$. Let $L_{b}$ be the left children of these nodes that do not lie on the path (see Figure~\ref{fig:search_path}). In $O(\log{n})$ time, the points of the nodes of $P_{b} \bigcup L_{b}$ that lie in the query-range can be determined. Then, for each node of $L_{b}$ storing a point inside the range query, its two children are visited and checked whether their points lie in the range. This procedure continues recursively, as long as points in the query-range are found. 

 The correctness of the query algorithm is proved as follows. First, observe that nodes to the right of the search path, have points with $x$-coordinate larger than $b$ and therefore lie outside the query-range. The points of $P_{b}$ may have $x$-coordinate larger than $b$ or they may have $y$-coordinate larger than $c$. In any case, they are not reported. The nodes of $L_{b}$ and their descendants have points with $x$-coordinate smaller than $b$, so that only their $y$-coordinates need to be tested. The children of nodes of $L_{b}$ with $y$-coordinate less than $c$ must be considered. In particular, the reporting procedure proceeds recursively, as long as points inside the query range are found. If a point of a node $u$ does not lie inside the query-range, then this point has $y$-coordinate larger than $c$. Therefore, all points in the subtree rooted at $u$ lie outside the query-range and they are not reported. We can easily bound the query time by $O(\log{n}+t)$, since $O(\log{n})$ time is needed to visit the nodes in $P_{b} \bigcup L_{b}$ and $O(t)$ time is necessary for the reporting procedure in their subtrees. 

\subsection{Persistent Modified Priority Search Tree}

 In this subsection, we briefly review the Modified Priority Search Tree of sioutas et al. \cite{SMNLTTV04}. 
Let $S$ be a set of $n$ points on the plane with coordinates $(x,y)$, where $x\in \{1,\ldots,M\}$ and $y \in \Re$. Without loss of generality we assume that all points are distinct. We will show how to store the points in a data structure $T$, so that the $t$ points in a query range of the form $\left(-\infty , b\right] \times \left(-\infty , c\right]$, can be found in $O(t)$ time. Our structure relies on the priority search tree, which we augment 
with list-structures similar to those in \cite{O88}.

We denote by $T_{v}$ the subtree of $T$ with root $v$.

The tree structure $T$ has the following properties:

\begin{itemize}
\item{Each point of $S$ is stored in a leaf of $T$ and the points are in sorted $x$-order from left to right.}
\item{Each internal node $v$ of $T$ stores a point $p(v)$ of $S$. The point $p(v)$ is the point with the minimum $y$-coordinate amongst the points stored in the leaves of $T_{v}$.}
\item{Each node $v$ is equipped with a secondary list $S(v)$. $S(v)$ contains the points stored in the leaves of $T_{v}$ in increasing $y$-coordinate.}
\end{itemize}

\begin{figure}[htbp]
	\begin{center}
		\includegraphics[scale=0.6]{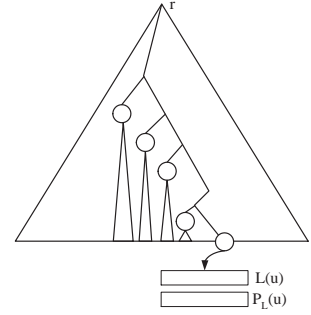}
\end{center}
 \caption{$L(u)$: it stores the points of the nodes of $L_{i}$. $P_{L}(u)$: it stores the points of the nodes of $P_{i}$ which have $x$-coordinate smaller or equal to $i$.}
 \label{fig:arrayslp}
\end{figure}

 For convenience we will assume that the tree $T$ is a complete binary tree (i.e. all its leaves have depth $\log{n}$). Note that if $n$ is not a power of $2$, then we may add some dummy leaves so that $T$ becomes complete. We also use an array $A$ of size $M$, which stores pointers to the leaves of $T$.  Specifically, $A[i]$ contains a pointer to the leaf of $T$ with maximum $x$-coordinate smaller or equal to $i$. This array is used to determine in $O(1)$ time the leaf of the search path $P_{b}$ for $b$. In each leaf $u$ of the tree with $x$-coordinate $i$ we store the lists $L(u)$ and $P_{L}(u)$. The list $L(u)$ stores the points of the nodes of $L_{i}$. The list $P_{L}(u)$ stores the points of the nodes of $P_{i}$ which have $x$-coordinate smaller or equal to $i$. Both lists also contain pointers to the nodes of $T$ that contain these points. Each list $L(u)$, $P_{L}(u)$, stores its nodes in increasing $y$-coordinate of their points (see Figure~\ref{fig:arrayslp}).

 To answer a query of the form $(-\infty , b] \times (-\infty , c]$ we find in $O(1)$ time the leaf $u$ of the search path $P_{b}$ for $b$. Then, we traverse the list $P_{L}(u)$ and report its points until we find a point with $y$-coordinate greater than $c$. We traverse the list $L(u)$ in the same manner and find the nodes $v$ of $L_{b}$ whose points have $y$-coordinate less than or equal to $c$. For each such node $v$ we traverse the secondary list $S(v)$ and report its points until we find a point with $y$-coordinate greater than $c$.
 
The following theorem bounds the size and the query time of our structure.

{\bf Theorem 1.}
Given a set of $n$ points on the plane with coordinates $(x,y)$ such that $x\in\{1,\ldots,M\}$ and $y\in \Re$, we can store them in a data structure with $O(M+n\log{n})$ space that supports quadrant range queries in $O(t)$ time, where $t$ is the number of reported points.

{\bf Proof.}
The query algorithm finds the $t'$ points of nodes of $P_{b} \bigcup L_{b}$ that lie inside the query-range in $O(t')$ time by simple traversals of the lists $P_{L}(u)$, $L(u)$. The search in the respective subtrees $T(v)$ can be performed by traversing the secondary lists $S(v)$ and takes $O(t)$ additional time for reporting $t$ points in total. Therefore, the query algorithm needs $O(t)$ time. Each list $P_{L}(u)$, $L(u)$ stores the respective points in the nodes of the path from root to $u$, and points in the left children of nodes of this path. So, the size of each list is $O(\log{n})$ and the space of $T$ is $O(n\log{n})$. The total space of secondary lists $S(v)$ is also $O(n)$. Thus, the space of the whole structure is $O(M+n\log{n})$ because of the size of the array $A$.

The $O(n\log n)$ term in the space bound is due to the size of the lists $P_{L}(u)$ and $L(u)$. We can reduce the total space of these lists to $O(n)$ by making them persistent. Ordinary structures are ephemeral in the sense that update operations make permanent changes to the structures. Therefore in ordinary structures it is impossible to access their old versions (before the updates). According to the terminology of Driscoll et al. \cite{DSST89} a structure is persistent, if it allows access to older versions of the structure. There are two types of persistent data structures: partially and fully persistent. A partially persistent data structure allows updates of its latest version only, while a fully persistent one allows updates of any of its versions. In \cite{DSST89}, a general methodology is proposed for making data structures of bounded in-degree persistent. With their method such a structure can be made partially persistent with $O(1)$ amortized space cost per change in the structure. In our case a list can be made partially persistent with a $O(1)$ amortized increase in space per insertion/deletion.
 
\begin{figure}[htbp]
	\begin{center}
		\includegraphics [scale=0.6]{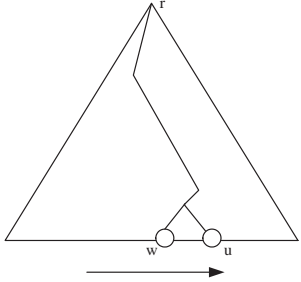}
 \end{center}
 \caption{Lists $P_{L}(u)$ and $L(u)$ are implemented as partially persistent lists, by performing a sweep from left to right.}
 \label{fig:sweep}
\end{figure}

 We show how to implement the lists $P_{L}(u)$ using a partially persistent list. Let $u$ be a leaf in $T$ and let $w$ be its predecessor (the leaf on the left of $u$). We denote by $x_{u}$ the $x$-coordinate of $u$ and by $x_{w}$, the $x$-coordinate of $w$. The two root-to-leaf paths $P_{x_{u}}$, $P_{x_{w}}$, share the nodes from the root of $T$ to the nearest common ancestor of $u$, $w$. As a result, we can create $P_{L}(u)$ by updating $P_{L}(w)$ in the following way. First we delete from list $P_{L}(w)$ the points that don't lie on $P_{x_{u}}$. Then we insert the points of $P_{x_{u}}$ which have $x$-coordinate smaller or equal to $x_{u}$. In this way we can construct all lists as versions of a persistent list: we begin from the leftmost leaf and construct the list $P_{L}(u)$ of each leaf $u$ by updating the one of its predecessor (see Figure~\ref{fig:sweep}). 

The total number of insertions and deletions is $O(n)$ because each point is inserted and deleted only once. Therefore the space of all the lists is $O(n)$. In the same way, lists $L(u)$ are constructed for all leaves in $O(n)$ space. Therefore:

{\bf Theorem 2.} 
Given a set of $n$ points on the plane with coordinates in the range $[1, M] \times R$  we can store them in a data structure with $O(n+M)$ space that allows quadrant range queries to be answered in $O(t)$ time, where $t$ is the number of answers.

The preprocessing time is $O(M+n\log n)$ but with a more careful implementation we can reduce this complexity to $O(M+n)$, by using the pruning technique as in \cite{FMNT}.

\subsection{Geometric Transformation Of Duality}

In homogeneous coordinates, there exists a duality between the point $(a,b,c)$ and the line $(ax+by+cz=0)$ for all $(a,b,c)\neq(0,0,0)$. Furthermore, the following pairs are dual:

\begin{itemize}
	\item points on a line $\longleftrightarrow$ lines throught a point
\end{itemize}

\begin{itemize}
	\item line segment $\longleftrightarrow$ "double wedge" of lines
\end{itemize}

\begin{figure}[htbp]
	\begin{center}
		\includegraphics [scale=0.6]{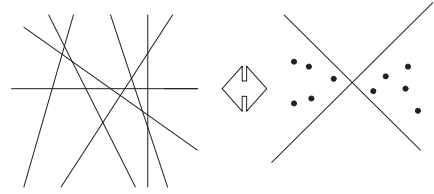}
	\end{center}
	\caption{An example of Geometric Duality Transformation: Dual pairs}
	\label{fig:Dual_pair}
\end{figure}

\begin{itemize}
	\item set of lines intersecting a line segment $\longleftrightarrow$ set of points in a double wedge (see Figure~\ref{fig:Dual_pair})
\end{itemize}

A more formal definition of this powerfull tool can be found in \cite{CGL83}.

\section{Algorithms for Orthogonal Objects} \label{sec:alg}

\subsection{Data Structures for Orthogonal Triangle Range Queries}

\begin{figure}[htbp]
	\begin{center}
		\includegraphics[scale=0.4]{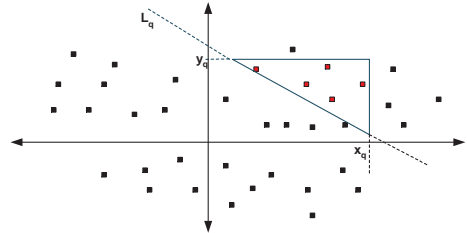}
	\end{center}
	\caption{An example of Orthogonal Triangle Range Query}
	\label{fig:triangle_query}
\end{figure}

 An orthogonal triangle range query is the problem of determining all the points from a set $S$ of $n$ points on the plane lying inside an orthogonal triangle. Recall, a triangle is orthogonal iff two of its edges are axis-parallel. Let $T$ be an orthogonal triangle defined by the point $(x_q,y_q)$ and the line $L_q$ that is not axis-parallel (see Figure ~\ref{fig:triangle_query}).

A retrieval query for this problem can be supported efficiently by the following data structures:

\begin{enumerate}

\item {\em{\bf First Solution: A 3-layered Tree:}} To set up the data structure, first sort the $n$ points according to their $x$-coordinates and then store the ordered sequence in a leaf-oriented balanced binary search tree of depth $O(\log{n})$. This structure answers the query: "determine the points having $x$-coordinates in the range $[x_1,x_2]$" by traversing the two paths to the leaves corresponding to $x_1$, $x_2$. The points stored as leaves at the subtrees of the nodes which lie between the two paths are exactly these points in the range $[x_1,x_2]$. For each subtree, the points stored at its leaves are organized further to a second level data structure according to their $y$-coordinates in the same way. For each subtree of the second level data structure, the points stored at its leaves are organized further to a third level modified data structure of Chazelle (presented in subsection 2.3) for half-plane range query.
So, each Orthogonal Triangle Range Query is performed through the following steps:

\begin{enumerate}

	\item In the tree storing the pointset $S$ according to $x$-coordinates, traverse the path to $x_q$. All the points having $x$-coordinate in the range $[x_q,\infty)$ are stored at the subtrees on the nodes that are right sons of a node of the search path and do not belong to the path. There are at most $\left\lceil \log{n} \right\rceil$ such disjoint subtrees.
	
	\item For every such subtree traverse the path to $y_q$. By a similar argument as in the previous step, at most $\left\lceil \log n \right\rceil$ disjoint subtrees are located, storing points that have $y$-coordinate in the range $[y_q,\infty)$.
	
	\item For each subtree in Step 2, apply the half-plane range query algorithm presented in subsection 2.3 in order to retrieve the points that lie on the side of line $L_q$ towards the triangle.  

\end{enumerate}

The correctness of the above algorithm follows from the data structure used. Since we have to visit $O(\log{n})$ subtrees in each of the two first layers and the third layer requires $O(\frac{logn}{loglogn}+A)$ time, the overall query time becomes $O(\frac{log^3{n}}{loglogn}+A)$ while the space complexity is $O(n \log^2{n})$.

\item {\em{\bf Second Solution: A two layered tree:}} The first layer is a linear space Persistent Modified Priority Search Tree. The second layer organizes each of the $O(\log^{2}n)$ subsets of $P_{L}(u)$ and $L(u)$ with the modified structure of \cite{CGL83} presented in subsection 2.3 which returns a set of nodes $v$ whose points lie inside the triangle. Then, it organizes each of the $O(n^{2})$ subsets of $S(v)$ with the same structure (see figure 6).

More specifically, we find in $O(1)$ time the leaf $u$ of the search path $P_{b}$ for $b$. Each leaf $u$ stores two y-ordered Persistent lists $P_{L}(u)$ and $L(u)$, both of size $O(\log{n})$. The total number of subsets, which can be produced by taking all the possible y-ordered permutations amongst the y-coordinates of $P_{L}(u)$ and $L(u)$, is $O(\log{n})$, since each list $P_{L}(u)$ or $L(u)$ consumes $O(\log{}n)$ space in worst-case. Each subset has minimum size $1$ and maximum $O(\log{n})$, thus their total space is $1+2+…+logn=O(\log^{2}n)$. We can access the appropriate subset of points (according to $y_{q}$ - coordinate of the query) in $O(\log\log{n})$ time by applying a simple binary searching. Then we apply in the subset above a half plane range query (according to query line $L_{q}$) in order to find the nodes $v$ whose points lie inside the triangle. For each such node $v$ we organize its secondary list $S(v)$ again as the modified structure of Chazelle presented in subsection 2.3. The total number of subsets, which can be produced by taking all the possible y-ordered permutations amongst the y-coordinates of $S(v)$, is $O(n)$, since each secondary list $S(v)$ consumes $O(n)$ space in worst-case. Each subset has minimum size $1$ and maximum $O(n)$, thus their total space is $1+2+…+n=O(n^{2})$. We can access the appropriate subset of points (according to $y_{q}$ - coordinate of the query) in $O(\frac{log{n}}{loglogn})$ time by using the fusion tree method \cite{FW90}. Then we apply in the subset above a half plane range query (according to query line $L_{q}$). As a consequence the overall query time becomes $O(logn/loglogn + A)$ and the space consumption is $O(n^{2})$. 

\begin{figure}[htbp]
	\begin{center}
		\includegraphics[scale=0.5]{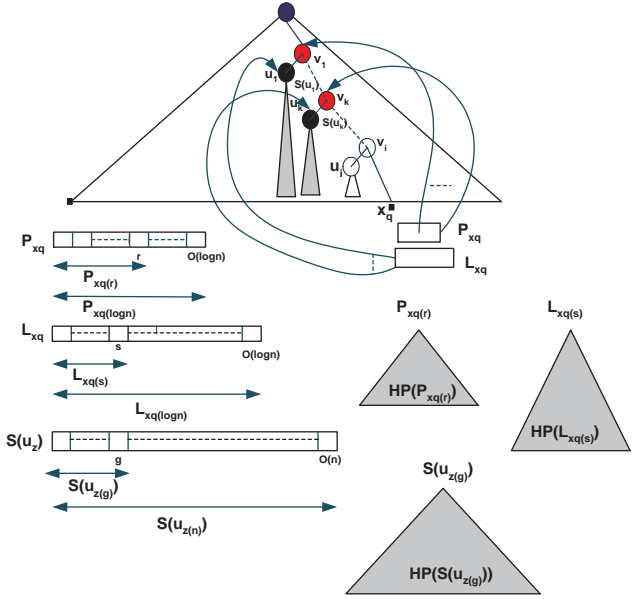}
	\end{center}
	\caption{The second solution. The lists P, L, S and its respective Half Plane (HP) structures}
	\label{fig:second_solution}
\end{figure}

\end{enumerate}

\subsection{The Results}

\begin{figure}[htbp]
	\begin{center}
		\includegraphics[scale=0.6]{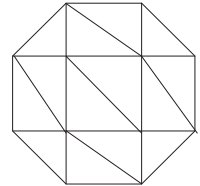}
	\end{center}
	\caption{Orthogonal triangle decomposition of a canonical octagon}
	\label{fig:octagon}
\end{figure}

 Every canonical $k$-vertex polygon can be decomposed into $O(k)$ orthogonal triangles because of the strict symmetry in the topology of the vertices. In Figure~\ref{fig:octagon} such a decomposition is depicted.
 
\begin{itemize}

	\item \emph{\textbf{Solution 1}}: Querying the 3-layered data structure $O(k)$ times, the $O(n \log^2{n})$ space and 
	$O(klog^3n/loglogn + A)$ time follows.

	\item \emph{\textbf{Solution 2}}: Querying the two layered data structure $O(k)$ times, the 
$O(n^2)$ space and $O(klogn/loglogn + A)$ time follows.

\end{itemize}

\section{Some Thoughts on the General Polygon Retrieval Problem} \label{sec:thoughts}

\subsection{The Algorithm of Paterson and Yao}

This section sketches the basic ideas of the algorithm for polygon retrieval on points that Paterson and Yao \cite{PY86} have presented.

This problem is to preprocess a set $S$ of $n$ points on the plane, so that for any query polygon $P$ the subset of them lying inside it can be reported efficiently. We assume that the given polygon is convex with $k$ vertices. If not, ($P$ is non-convex) then a suitable decomposition into convex parts can be carried out and the algorithm can be applied to every such part. The key-idea of Paterson/Yao's algorithm is a further decomposition of every convex part into $O(k)$ simpler parts called \emph{quads}. They solve the problem for each such quad separately, using the well-known geometric transformation of duality. 

To set up the data structure, Paterson/Yao first sort the $n$ points according to their polar angles at the origin and then store the ordered sequence in a leaf-oriented balanced binary search tree of depth $O(\log{n})$. This structure answers the query: "determine the points having polar angle in the range $[a_1,a_2]$ by traversing the two paths to the leaves corresponding to $a_1$, $a_2$. The points stored as leaves at the subtrees of the nodes which lie between the two paths are exactly these points in the range $[a_1,a_2]$. For each subtree, the points stored at its leaves are organized further to a second level data structure as follows: by the duality correspondence, these points are mapped to a set of straight lines, and a point location structure is built for this planar subdivision \cite{EGS84}. Using this data structure a query of the form: "Report the points lying in a double wedge" is reduced to a query to the dual structure of the form: "report the lines that intersect a query line segment."

 At this point, we are ready to give a step by step description of the polygon retrieval algorithm of Paterson and Yao: 
 
\begin{enumerate}
	
	\item Let $P$ be the $k$-vertex convex query polygon. Separate $P$ into simpler regions called \emph{quads} by cutting the plane into \emph{sectors}. The \emph{sectors} are obtained by drawing semi-infinite lines from the \emph{origin} to each vertex of $P$. Each \emph{quad} has a particularly simple form (it is the intersection of a sector and a double wedge). (see Figure ~\ref{fig:sectors_and_quads})

\begin{figure}[htbp]
	\begin{center}
		\includegraphics[scale=0.4]{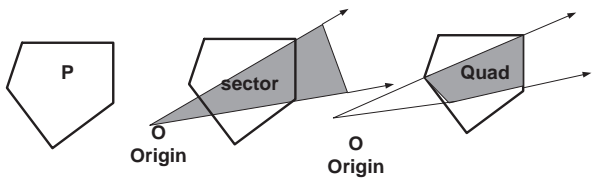}
	\end{center}
	\caption{An example of Sectors and Quads}
	\label{fig:sectors_and_quads}
\end{figure}

	Every $k$-gon is the disjoint union of $O(k)$ such \emph{quads}.
	
	\item Using the two-level data structure do the following for each quad with bounding rays $\ell_1$, $\ell_2$:

\begin{itemize}
	
	\item Determine the (at most) $2\left\lceil logn\right\rceil$ disjoint subtrees that descend from the two search paths to the angles of $\ell_1$, $\ell_2$ and lie between them. All the points between $\ell_1$ and $\ell_2$ are stored as leaves in these subtrees.
	
	\item for each such subtree in turn report the points inside the double wedge determining the quad using the dual point location structure.
	
\end{itemize}
	
\end{enumerate}

The space used is $O(n^2)$ because of the point location structure and the query time is $O(k\log{n}+A)$ (there exist $O(k)$ quads and the time spent for every "quad-retrieval" is $O(\log^2{n}+A_i)$ which can be improved using the fractional cascading technique \cite{CG85} on the point location structures). It is worthwile to note that the above algorithm works even when $P$ is unbounded. If $P$ is non-convex, then the preliminary decomposition step into convex parts does not change the asymptotic space and time bounds. 

Unfortunately, the $O(n^2)$ space bound is the best possible if we directly use the dual point location structures. This means that in order to reduce the space requirements of the later algorithm, we have to find an algorithm supporting "quad retrieval" without using the duality on arbitrary sets of $O(n)$ points which lies to $O(n^2)$ space subdivisions. The next section demonstrates such an algorithm working efficiently when $P$ is a canonical $k$-vertex polygon. 

\subsection{An Extended Approach}

\begin{figure}[htbp]
	\begin{center}
		\includegraphics[scale=0.7]{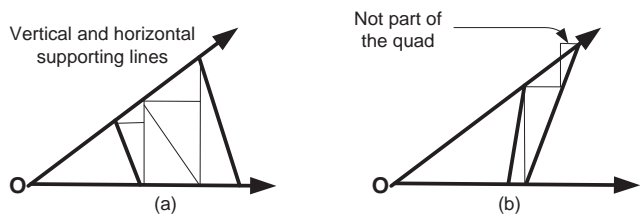}
	\end{center}
	\caption{An example of Quad Decomposition}
	\label{fig:Decomposition_of_quad}
\end{figure}

In this section we present an extended approach for the general polygon retrieval problem. The proposed solution is based on the algorithm of Paterson/Yao for answering "quad retrieval" queries, which was described in the previous section. The key-idea is the reduction of a quad retrieval query to a constant $O(1)$ number of orthogonal triangles retrieval queries. This is achieved by drawing parallel lines (vertical or horizontal) from the vertices of the quad so that the lines stay inside the quad and intersect the rays from the origin. To do this, a suitable decomposition of each quad is needed. In the general case of an arbitrary (non canonical) $k$-vertex polygon, such a decomposition is not always possible. For example, the quad in Figure~\ref{fig:Decomposition_of_quad}(a) can be decomposed into $O(1)$ orthogonal triangles while the quad in Figure~\ref{fig:Decomposition_of_quad}(b) cannot be decomposed.
 
 This of course depends on various geometric characteristics of the quad. If the polygon is $k$-vertex canonical, we can always directly carried out a decomposition into $O(k)$ orthogonal triangles because of the strict symmetry in the topology of the vertices. Arbitrary $k$-vertex polygons satisfy such a decomposition only in special cases.

\section{Conclusions} \label{sec:concl}

 In this work we presented efficient algorithms for the problem of canonical polygon retrieval queries on the plane. One deficiency or our algorithms is that they are static. It seems very interesting the dynamization of presented algorithms. Furthermore, it still remains an open problem whether an $O(n\log^{O(1)}{n})$ space and $O(\log^{O(1)}n+A)$ time algorithm exists for the general problem of arbitrary polygon retrieval queries. Finally, it would be of great importance the incorporation of the proposed algorithms in \emph{meandric polygons} \cite{PV08}, a specific kind of planar polygons with many applications in bioinformatics. The latter constitutes a real case study in bioinformatics, which we plan to implement in the near future.

\end{document}